\documentclass[namedreferences]{solarphysics}
\usepackage[optionalrh]{spr-sola-addons} 
\usepackage{graphicx}        
\usepackage{color}           
\usepackage{url}        

\newcommand{\kms}{km~s$^{-1}$ }

\chardef\us=`\_

\begin{document}

\begin{article}
\begin{opening}

\title{Development of a Confined Circular-cum-parallel Ribbon Flare and Associated Pre-flare Activity}
\author[addressref={aff1},corref,email={setiapooja.ps@gmail.com}]{\inits{Pooja}\fnm{Pooja}~\lnm{Devi}}
\author[addressref=aff2]{\inits{Bhuwan}\fnm{Bhuwan}~\lnm{Joshi}}
\author[addressref=aff1]{\inits{Ramesh}\fnm{Ramesh}~\lnm{Chandra}}
\author[addressref=aff2]{\inits{Prabir}\fnm{Prabir K.}~\lnm{Mitra}}
\author[addressref=aff3]{\inits{Astrid}\fnm{Astrid M.}~\lnm{Veronig}}
\author[addressref=aff1]{\inits{Reetika}\fnm{Reetika}~\lnm{Joshi}}

\address[id=aff1]{Department of Physics, DSB Campus, Kumaun University, Nainital 263002, India}
\address[id=aff2]{Udaipur Solar Observatory, Physical Research Laboratory, Udaipur 313004, India}
\address[id=aff3]{Institute of Physics $\&$ Kanzelh\"ohe Observatory for Solar and Environmental Research, University of Graz, A-8010 Graz, Austria}

\runningauthor{P. Devi {\it et al.}}
\runningtitle{Confined Circular Ribbon Flare}

\begin{abstract}
We study a complex GOES M1.1 circular
ribbon flare and related pre-flare activity
on 26 January 2015 [SOL2015-01-26T16:53] in solar active region NOAA 12268.
This flare activity was observed by
the  {\it Atmospheric Imaging Assembly} (AIA) on board {\it Solar Dynamics Observatory} (SDO)
and the {\it Reuven Ramaty High Energy Solar Spectroscopic Imager} (RHESSI).
The examination of photospheric magnetograms during the extended period, prior to the event,
suggests the successive development of a so-called $``$anemone$"$ type magnetic configuration.
The Nonlinear Force Free Field (NLFFF) extrapolation reveals a fan-spine magnetic configuration with the presence
of a coronal null-point. 
We found that the pre-flare activity in the active region starts $\approx$15 min
prior to the main flare in the form of localized bright patches at two locations.
A comparison of locations and spatial structures of the pre-flare activity with magnetic configuration
of the corresponding region suggests onset of magnetic reconnection at the null-point along with the
low-atmosphere magnetic reconnection caused by the emergence and the cancellation of the magnetic flux.
The main flare of M1.1 class is characterized
by the formation of a well-developed circular ribbon along with a region of remote brightening.
Remarkably, a set of relatively compact parallel ribbons formed inside the periphery of the
circular ribbon which developed lateral to the brightest part of the circular ribbon.
During the peak phase of the flare, a coronal jet is observed at the north-east
edge of the circular ribbon which suggests interchange reconnection between large-scale
field lines and low-lying closed field lines. Our investigation suggests a combination of 
two distinct processes in which ongoing pre-flare null-point reconnection
gets further intensified as 
the confined eruption along with jet activity proceeded 
from within the circular ribbon region which results to the 
formation of inner parallel ribbons and corresponding post-reconnection arcade.
\end{abstract}

\end{opening}

\section{Introduction}
\label{S-Intro}
Solar flares are the result of a sudden release of free magnetic energy stored in active regions 
\citep{Parker63,Fletcher11,Shibata11}.
However, there are also the cases where flares can occur outside the active region (for example \opencite{Chandra17}).
Broadly they can be classified in two categories \emph{i.e.} eruptive and confined flares
\citep{Svestka1986}. Eruptive flares are usually
long duration two-ribbon flares 
that show prolonged energy release (up to several hours).
These type of flares occur in association
with coronal mass ejections (CMEs).
In contrary, confined flares are short duration flares showing localized brightening
and are not accompanied by CMEs
(for detail see the
review by \opencite{Shibata11}).\

According to the morphology, a new category of flares have been reported which are known as circular ribbon flares \citep{Wang12}.
These flares usually contain a circular or quasi-circular ribbons and an inner ribbon inside it.
Circular ribbon flares can be confined \citep{Masson09,Zuccarello17,Hernandez-Perez2017} as well as eruptive in nature \citep{Hong17,Li18}.
According to the magnetic topology, circular ribbon flares occur
in anemone active regions \citep{Asai08}.
Anemone active regions are specified as active regions, where a single polarity is surrounded by opposite polarity in a circular pattern.
In some of the circular ribbon flares, one remote ribbon is
 present and it normally appears after
the formation of the circular ribbon.
However, there are also cases where the remote 
ribbon appears simultaneously with the circular ribbon.
Another important ingredient, which has been observed in many circular ribbon flares, is the presence of jets/surges at the boundary of the circular ribbon. 
The combination of circular and remote ribbons along with jets/surges
 were also studied in the numerical simulations
 \citep{Masson09,Pariat10}. According to the literature,
the first detailed study of a circular ribbon flare was presented by
\inlinecite{Masson09}. Since then circular ribbon flares have received quite
some attention which resulted in many subsequent studies
({\it e.g.,} \opencite{Reid12,Wang12,Sun13,Jiang13,Xu17,Hernandez-Perez2017}).
Very recently, \inlinecite{Hao17} reported white light kernels in a circular ribbon flare.
Further, with high resolution IRIS data, the characteristics of two circular ribbon flares were studied \citep{Li18}.

To explain the general characteristics of eruptive solar flares including the evolution of flare ribbons and loops, the standard CSHKP
model was put forward by \cite{Carmichael64}, 
\cite{Sturrock66}, \cite{Hirayama74}, and \cite{Kopp76}. However, this 2-Dimensional model does not
 explain the formation of circular flare ribbons.
 Recently, \inlinecite{Aulanier12} extended the 2D model into 3D (see also the review by \opencite{Janvier15}).
Circular ribbon flares provide us an opportunity to understand the flare in 3D
because they are related to magnetic field topology of fan-spine or 3D null-point magnetic topology in the corona.
Although many circular ribbon flares have been reported, there are still several questions about the origin of such flares, such as,
why only some flares have remote ribbons together with circular ribbon?;
What is the temporal relation between circular and remote ribbons?;
Why are jets/surges not always present in all the cases?;
Why only few anemone active region produce such type of flares?; etc.

With reference to the above questions, we present here a study of a circular ribbon 
flare of GOES class M1.1 observed on 26 January 2015 using multi-instrument data.
An important aspect of present observations is the identification of distinct pre-flare activity which
initiates $\approx$15 min prior to the circular ribbon flare.
The flaring site exhibits classical fan-spine configuration with a coronal null-point while the underlying
photospheric fields shows $``$anemone$"$ type morphology at the site of fan-dome.
The main phase of the flare is characterized by the formation of circular ribbon and subsequent
remote ribbon with fine-structured kernels. Afterwards, the circular ribbon evolves into the parallel
ribbons with the formation of a post-flare loop between a set of conjugate (parallel) ribbons.
Further important finding includes the apparent progression of brightness at remote ribbon
location which we attribute to the slipping reconnection.
The paper is organized as follows: Section 2 provides the information of the data sets used for the investigation.
The analysis and results of the event based on EUV/UV and X-ray observations are given in Section 3.
Finally, we discuss and conclude our results in Section 4.

\section{Observational Data Sets}
For the current study, we used the data from following sources:
\\

\begin{itemize}

\item\textbf{SDO/AIA Data:} 
The \emph{Atmospheric Imaging Assembly\/} (AIA: \opencite{Lemen12})
on board the \emph{Solar Dynamics Observatory\/} (SDO: \opencite{Pesnell12}) 
observes the different layers of the Sun in seven EUV (94 \AA\ (6 MK), 131 \AA\ (10
MK), 171 \AA\ (600,000 K), 193 \AA\ (1 MK), 211 \AA\ (2 MK), 304 \AA\ (50,000 K),
and 335 \AA\ (2.5 MK)), two UV (1600 \AA\ (10,000 K), and 1700 \AA\ (4500 K)), and one
White-light (4500 \AA\ (6000 K)) channels with a cadence of 12 second, 24
second, and 3600 second, respectively. The pixel size of AIA data is 0.6$''$.
In this paper, we have analyzed AIA images
in 94 \AA, 171 \AA, 304 \AA, and 1600 \AA\ wavelengths to study the flare
in the corona, upper transition region, chromosphere, and photosphere, respectively.

\medskip

\item\textbf{SDO/HMI Data:} 
The \emph{Helioseismic Magnetic Imager\/}  (HMI: \opencite{Schou12}) on board
 SDO observe the full disk of the Sun at 6173 \AA\ with a spatial resolution of 1$''$.
 Here, we use LOS magnetic field maps, vector magnetograms,
and white light data of HMI to study the magnetic structure and
sunspot distribution of the active region, respectively.
\medskip

\item\textbf{RHESSI Data:}
The \emph{Reuven Ramaty High Energy Solar Spectroscopic Imager\/}
(RHESSI: \opencite{Lin02}) performs imaging and spectroscopy
of X-ray and gamma-rays emitted during solar flares.
RHESSI observes the full disk of the Sun with a spatial resolution as fine
as 2.3$''$ (energy range from 3 keV to 20 MeV).
We reconstructed RHESSI imaging using the {\it CLEAN} algorithm.
We have used all 9 front detector segments except 2 and 7 to reconstruct the images.
The spatial sampling of the reconstructed images is 2.0$''$ per pixel.
\medskip

\item\textbf{GONG Data:} 
The \emph {Global Oscillation Network Group\/} (GONG: \opencite{Harvey11})
provides full disk H$\alpha$ filtergram of the Sun.
The temporal and spatial resolutions are 1 min and 2$''$, respectively.
\end{itemize}

\section{Analysis and Results}
Active region NOAA 12268 emerged on the solar disk on 23 January 2015
at the heliographic location S10E67 with a $\beta$ type of magnetic configuration.
As time progressed, the magnetic configuration of the active region changed to more complex $\beta\gamma$ type. 
The active region completed its journey on
the solar disk on 04 February 2015 and its magnetic configuration 
remained $\beta\gamma$.

\begin{figure*}[ht!]
\centering
\includegraphics[width=\textwidth]{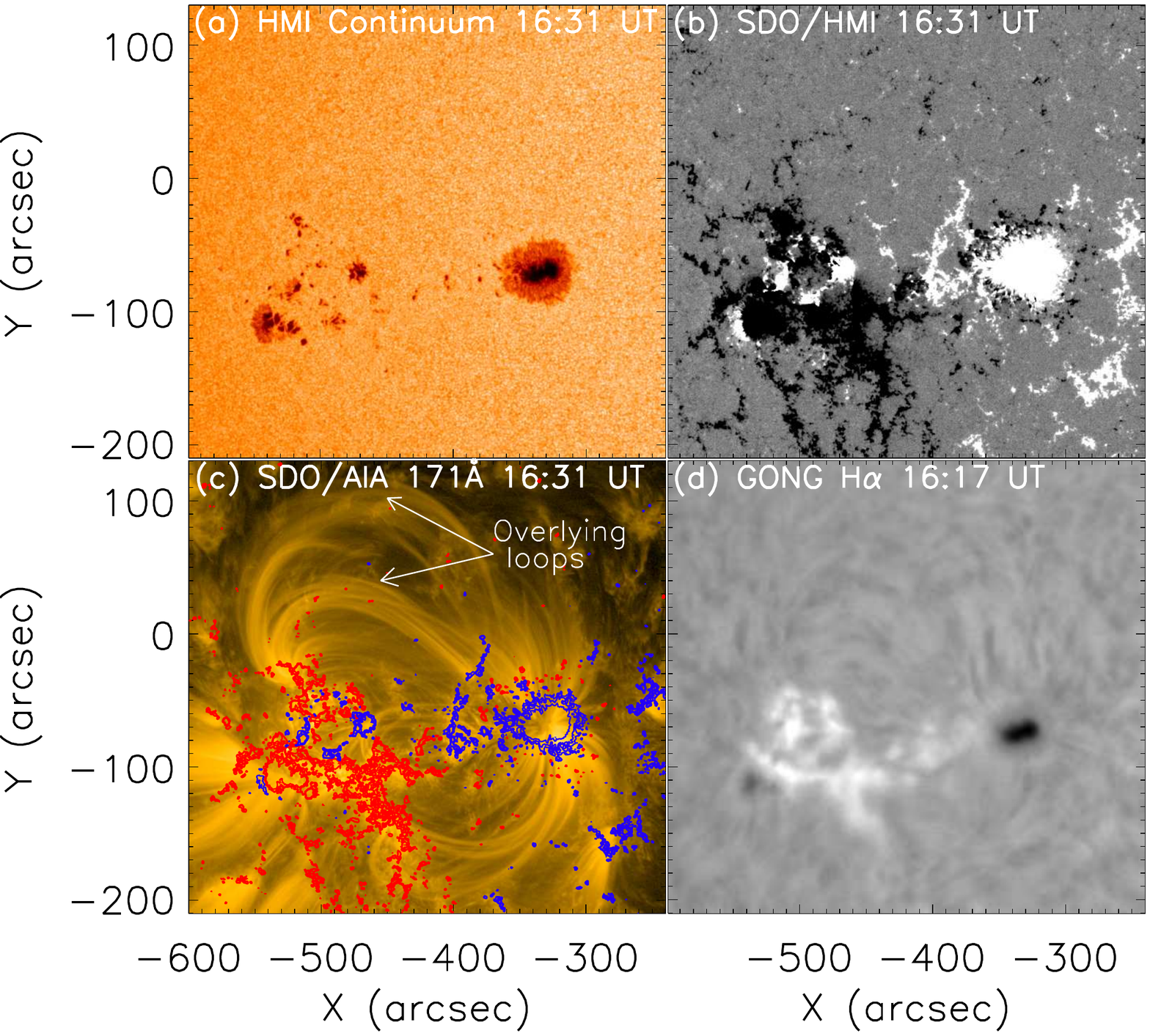}
\caption{Multi-wavelength view of NOAA AR 12268.
HMI continuum (a) and HMI magnetogram (b) showing that the
leading sunspot have positive magnetic polarity
and multiple smaller trailing sunspots have mixed (positive and negative) magnetic polarity. 
The AIA 171 \AA\ image with over-plotted magnetic field is shown in (c). 
Blue and red contours corresponds to positive and 
negative magnetic polarities, respectively.
The levels of these contours are $\pm$150, $\pm$300, and $\pm600$ G, respectively. 
The GONG H$\alpha$ image is shown in (d).}
\vspace{-6cm}
\label{Fig_overview}
\end{figure*}
On 26 January 2015, the active region produced a GOES flares of class
M1.1 (peak time $\approx$16:53 UT).
On that day, the active region was located at S10E35 on the solar disk.
The overview of the active region observed by SDO in different channels and
H$\alpha$ by GONG is presented in Figure $\ref{Fig_overview}$.
The active region consisted of a large leading sunspot of positive polarity followed by trailing sunspot groups of mixed polarity region with predominantly negative polarity. Since the flares occurred in the following polarity region, the location of the main flare was at the region
where positive magnetic polarity was surrounded by negative magnetic polarity.
The active region was also very active during 29--30 January 2015 and produced six flares of moderate intensities \citep{Zhong18}.

\begin{figure*}[ht!]
\centering
\includegraphics[width=\textwidth]{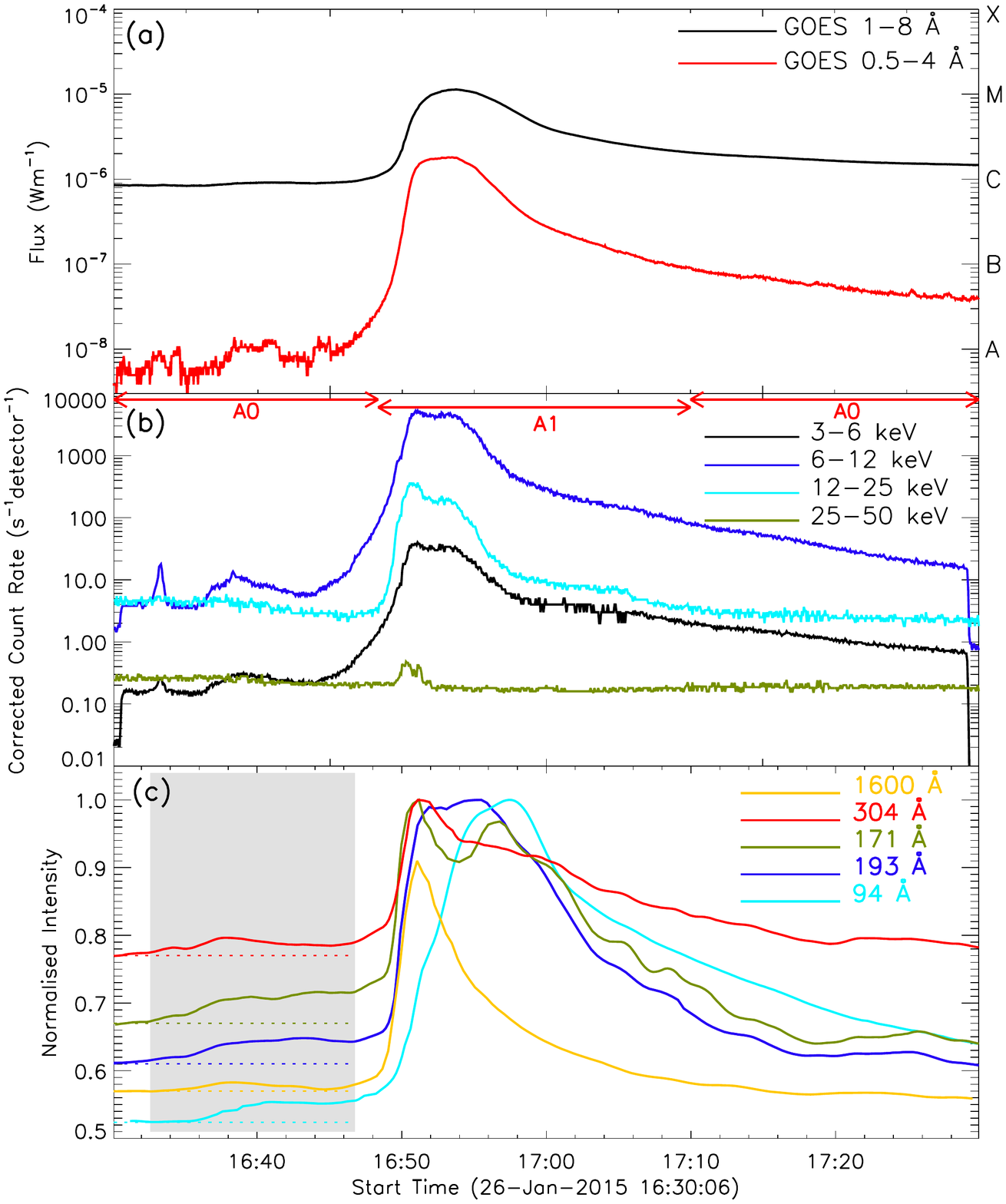}
\caption{GOES, RHESSI and AIA light curves showing 
the evolution of the M1.1 class flare.
The black and red curves in panel (a)
show the GOES X-ray flux in the 1--8 \AA\ and 0.5--4 \AA\ wavelength bands, respectively. 
Panel (b) shows RHESSI light curves for energy bands of 
3-6 keV (black), 6-12 keV (blue), 12-25 keV (sky), and 25-50 keV (green).
A0 and A1 denote the RHESSI attenuator states of the RHESSI.
Panel (c) shows the light curves of AIA for 94 \AA\ (sky), 193 \AA\ (blue), 
171 \AA\ (green), 304 \AA\ (red), and 1600 \AA\ (yellow).
The integrated intensity of different
channels is normalized by the maximum of corresponding light curve.
The grey shaded region corresponds to the pre-flare activity in the same active region.
For better visualization, the RHESSI count rates in the 6--12 and 
25--50 keV energy bands are
scaled by 10 and 1/100, respectively and the normalized intensity corresponding to the AIA UV
(1600 \AA) wavelength is scaled by a factor of 0.9 of its maximum  intensity.}
\vspace{-2.5cm}
\label{goes_rhessi_aia}
\end{figure*}
According to the GOES X-ray observation, the flare started $\approx$16:46 UT,
peaked at $\approx$16:53 UT and ended at $\approx$16:58 UT.
The flare was compact in nature and not associated with a filament eruption or CME.
However, during the flare, jet activity was observed, suggesting a confined eruption
along large-scale field lines.
Figure \ref{goes_rhessi_aia} displays the temporal evolution of the flare in X-rays and EUV/UV wavelengths.
The time profiles can be divided into four parts namely, pre-flare phase, impulsive phase, maximum phase, and  decay phase.
The description of these phases are presented as follows:

The pre-flare phase (between 16:32--16:48 UT) is visible in the GOES 0.5--4 \AA\ channel of
and RHESSI 3--6 and 6--12 keV energy bands along with several AIA EUV/UV (1600, 304, 193, 171, and 94 \AA)
wavelengths (see Figure \ref{goes_rhessi_aia}). 
The impulsive phase (between 16:48--16:50 UT) of the flare starts $\approx$4 min earlier in
the RHESSI lower energy bands (3--6 and 6--12 keV) in comparison to higher energy bands
(12--25 and 25--50 keV). The maximum phase (starting from $\approx$16:50 UT)
shows two peaks at 16:51 UT and 16:58 UT in all the RHESSI energy bands
and in some of the AIA EUV channels (193 and 171 \AA).
Here it is to be noted that AIA 94, 304, and 1600 \AA\ shows only one peak.
The peak of 304 and 1600 \AA\ are co-temporal with the first peak, whereas
AIA 94 \AA\ coincides with the second peak time.
After maximum, the decay phase of the flare starts from $\approx$17:00 UT onward,
showing a gradual decrease.

\subsection{Pre-flare Activity}
The pre-flare phase is defined between 16:32 UT and 16:48 UT.
We found slight increase in the intensity of the flaring region during this period. Figures \ref{Fig_mosac304}a and \ref{Fig_mosac304}b show the evolution of the pre-flare activity in AIA 304 \AA\ images. The first brightening in the active region was observed around 16:34 UT. After $\approx$3 minutes, the brightening becomes more significant (see Figure \ref{Fig_mosac304}b). Notably, pre-flare brightening at $\approx$16:37 UT agrees with the intensity enhancement seen in the AIA 304 \AA\ time profile of the circular ribbon region shown in Figure \ref{Fig_304light_curve}b.
On the other hand, the pre-flare brightening in the hot AIA 94 \AA\ filter
(temperature $\approx$6 MK) or
higher energy begins about $\approx$6 min later ($\approx$16:40 UT).
The pre-flare image in AIA 94 \AA\ is displayed in Figure \ref{Fig_mosac94}a which confirms the
location of the pre-flare brightenings to be co-spatial with the brightest emitting region of the circular ribbon during the peak phase of the flare.

As discussed above, the pre-flare phase was observed in RHESSI measurements
up to 12 keV energies. For the whole pre-flare phase, the RHEESI detectors were in the
{\it A0} attenuator state, {\it i.e.}, the imaging was done
with the maximum sensitivity limit of the detectors.
Therefore, the X-ray images during the pre-flare phase with mild emission are quite reliable.
To discuss the spatial evolution of the X-ray sources, we have reconstructed pre-flare
images in 3--6 and 6--12 keV energy bands (Figure \ref{Fig_rhessi_evolution}a and \ref{Fig_rhessi_evolution}b).
We found single and extended X-ray sources in both energy bands ({\it i.e.} 3--6 and 6--12 keV)
and the centroid of the sources were almost co-spatial. The location of the X-ray sources
were consistent with the EUV brightening. The pre-flare activity suggests that the
region where the circular ribbon developed revealed already energy release in the
pre-flare phase.
This shows the connection between the pre-flare activity and the later main flare reconnection. 

\subsection{Development of Circular Ribbon Flare and Remote Brightening}
\begin{figure*}[ht!]
\includegraphics[width=\textwidth]{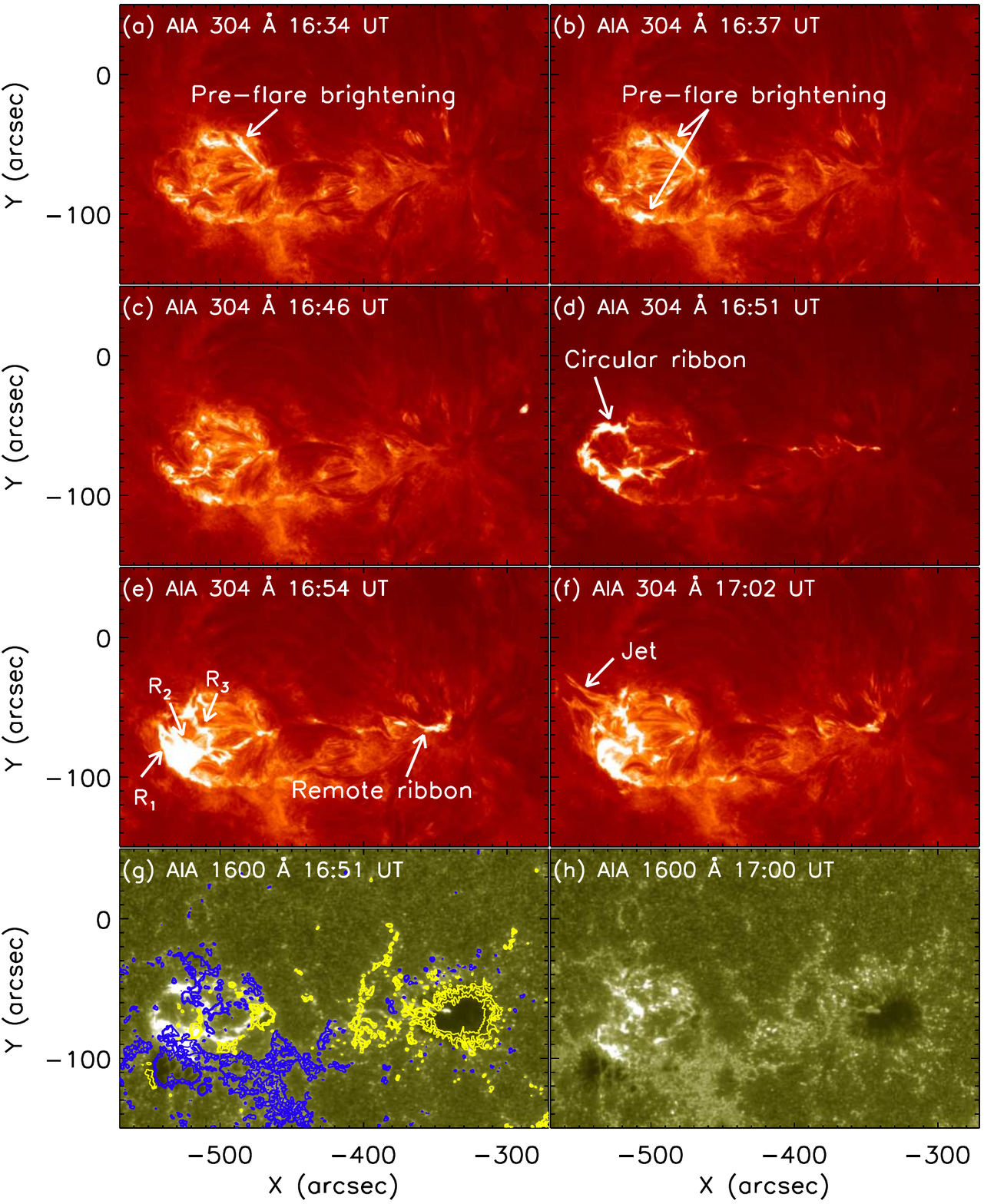}
\caption{Sequence of AIA 304 \AA\ images showing snapshots during the
pre-flare, main flare and post-flare phases:
(a) and (b) show the pre-flare images during the times of enhanced emission in GOES,
RHESSI, and AIA light curves, (c) corresponds to the impulsive phase of the flare.
Image (d) corresponds to the peak time 
of the event with the circular ribbon indicated by an arrow.
The remote brightening, flare ribbons (R1, R2, and R3) and jet structures 
are shown in (e) and (f), respectively.
Panels (g) and (h) show the images in AIA 1600 \AA\ 
wavelength corresponding to the peak phase of the flare.
The yellow and blue contours corresponds to the
 positive and negative magnetic fields, respectively.}
\label{Fig_mosac304}
\vspace{-2cm}
\end{figure*}

\begin{figure*}
\centering
\includegraphics[width=\textwidth]{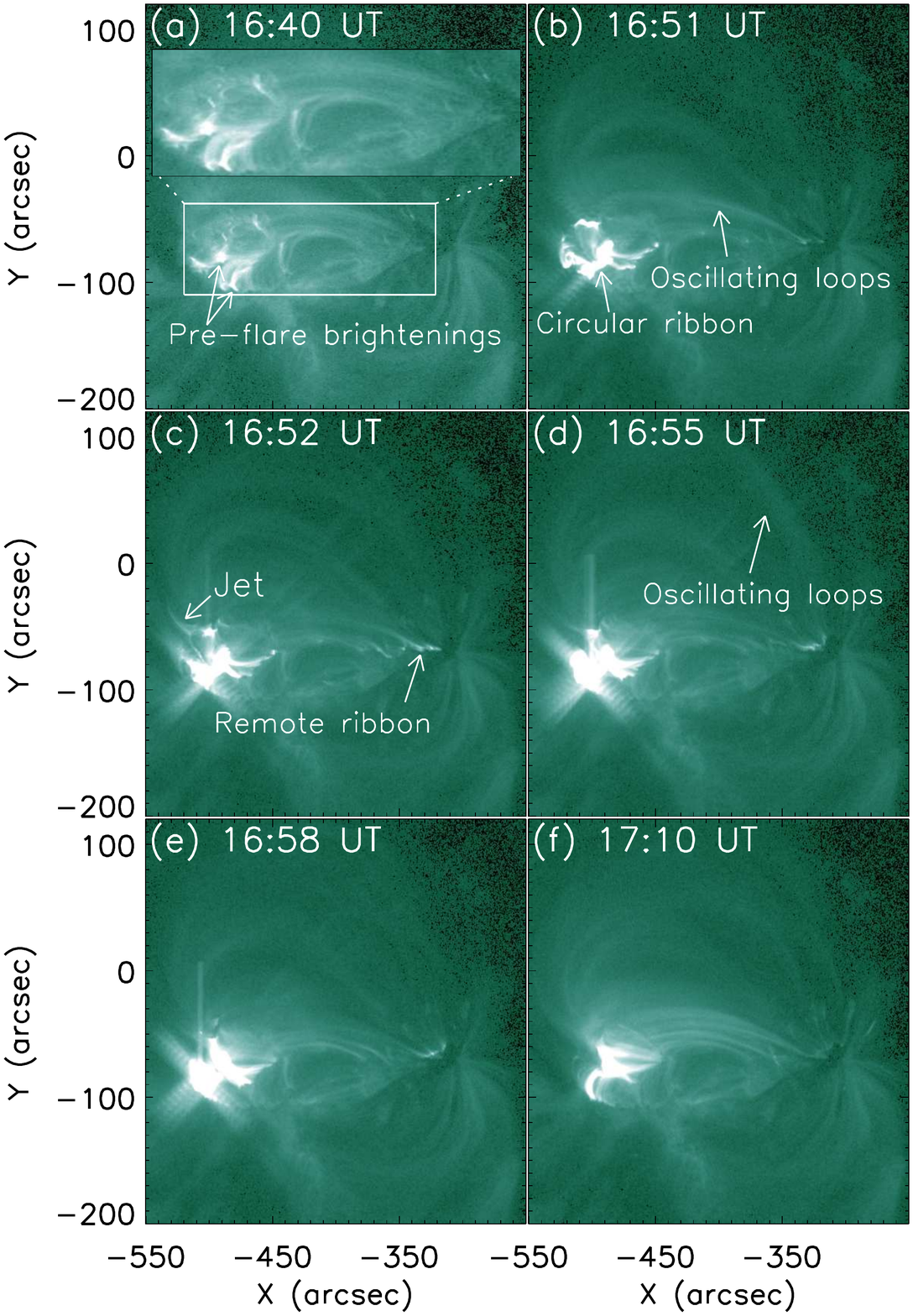}
\caption{Sequence of AIA 94 \AA\ images. 
Panel (a) shows pre-flare brightening as marked by arrows. The zoomed view of in the 
inset of panel (a) shows
the pre-flare brightening and the loops connecting the flare region and remote brightening region.
In image (b), the circular flare ribbon is shown. 
We can see the oscillation loops after the reconnection as shown in (b) and (d).
The jet and remote brightening associated with the flare are shown in (c).
The decay phase of the flare with large-scale hot loops connecting the main and the remote ribbons is shown by (e) and (f). }
\vspace{-0.3cm}
\label{Fig_mosac94}
\end{figure*}

From the onset of the impulsive phase, some faint circular ribbon flare structure started to appear.
As time progressed, the brightening spread in a circular 
path from [-500$''$,-85$''$] to [-525$''$,-65$''$] in clockwise
direction and finally the circular ribbon developed 
completely during the peak phase of the flare ($\approx$16:51 UT).
The development of the circular flare ribbon is presented 
in Figure \ref{Fig_mosac304}.
Importantly, at higher coronal temperatures (AIA 94 \AA\ images) also, the circular ribbon became visible around 16:51 UT (Figure \ref{Fig_mosac94}b).
However, the intensity of the flaring region in 94 \AA\ channel shows a delayed peak around 16:58 
UT as the intensity is dominated by loop brightening over the ribbon structure (see also the accompanying movie in AIA 94 and 304 \AA).
To investigate the magnetic field configuration 
at the flare site, we have overlaid the contours 
of the HMI line-of-sight magnetic field in AIA 1600 \AA\
image which is displayed in Figure \ref{Fig_mosac304}g.
From this figure, we can see that the circular ribbon developed along the circular polarity inversion line.

At $\approx$16:51 UT, bright material started to eject along the north-east direction from the brightest south-west part of the circular ribbon and formed ribbon R$_2$. Now looking at the flare morphology around 16:52 UT, we recognize three flare ribbons, R$_1$, R$_2$, and R$_3$ (shown by arrows in Figure \ref{Fig_mosac304}e). Parallel ribbon formation during a circular
ribbon flare was also observed by \inlinecite{Zhong18}
from the same active region during 29--30 January 2015.
Our flare and the flare studied by \inlinecite{Zhong18} bear some similarity as well as differences.
\cite{Zhong18} observed that the parallel ribbons converted
from the circular ribbon in the west part of the circular ribbon.
However, our parallel ribbons are not situated at the location found in the event of \inlinecite{Zhong18}.
In our case the conjugate parallel ribbons (R$_1$ and R$_2$)
form at the eastern part of the circular ribbon.
The third ribbon (R$_3$) is situated at the western part of the circular ribbon.
We have also noticed the two ribbon R$_1$ and R$_2$ 
separated as the time progresses as commonly observed in the
typical two ribbon flares. We could also see clearly the loops connecting these two ribbons.

\begin{figure*}
\centering
\includegraphics[width=\textwidth]{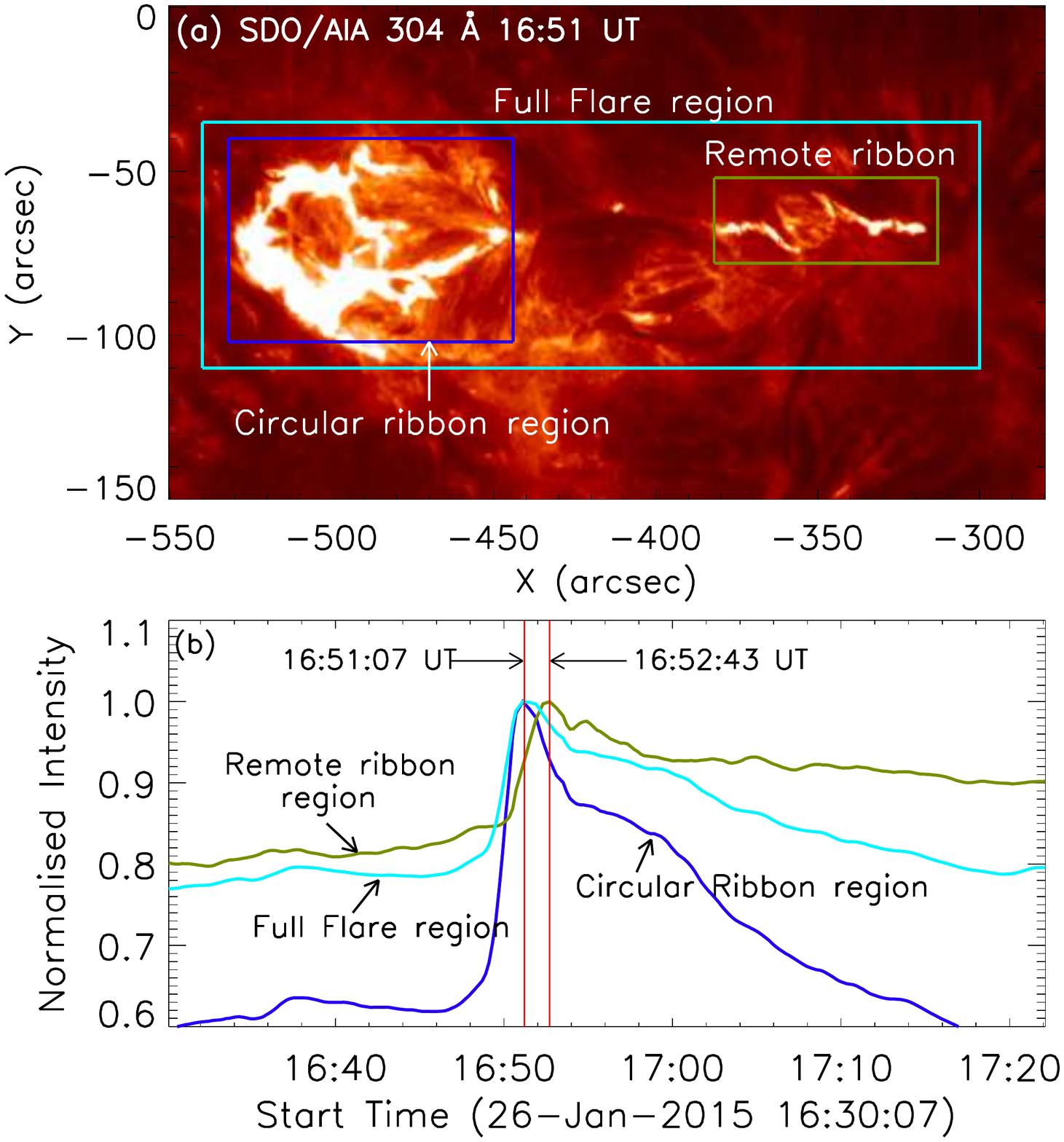}
\caption{AIA~304~\AA~image at the flare peak time (16:51 UT) and light curves for the selected regions.
Panel (a) shows the flare region where the boxes of 
cyan, blue, and green colors represent the full flare region, circular ribbon region, and remote brightening region, respectively.
The light curves corresponding to these boxes are shown in panel (b). 
Two vertical red lines in panel (b) show the peak time of the flux for these regions. 
The observed intensity of each curve is normalized by the corresponding peak intensity.}
\vspace{-4.1cm}
\label{Fig_304light_curve}
\end{figure*}

After $\approx$2 minutes of the 
complete development of the circular ribbon a remote brightening appears at the location [-390$''$,-70$''$],
$\approx$130$''$ west from the middle of the circular flaring region.
This remote brightening apparently slides 50$''$ westward after its first appearance
and eventually manifested in terms of a well developed remote ribbon.
The remote ribbon are displayed in Figures \ref{Fig_mosac304}e, \ref{Fig_304light_curve}a and
\ref{Fig_mosac94}c.
From the time profile of region encompassing the remote ribbon, we found that the
intensity of remote ribbon peaked $\approx$16:53 UT (Figure \ref{Fig_304light_curve}b).
The connectivity of the circular ribbon and remote ribbon can be seen in
Figure \ref{Fig_mosac94} through connecting loops
(marked by arrow in Figure \ref{Fig_mosac94}d).
We also observed the oscillation of these loops during the flare.

\begin{figure*}
\centering
\includegraphics[width=\textwidth]{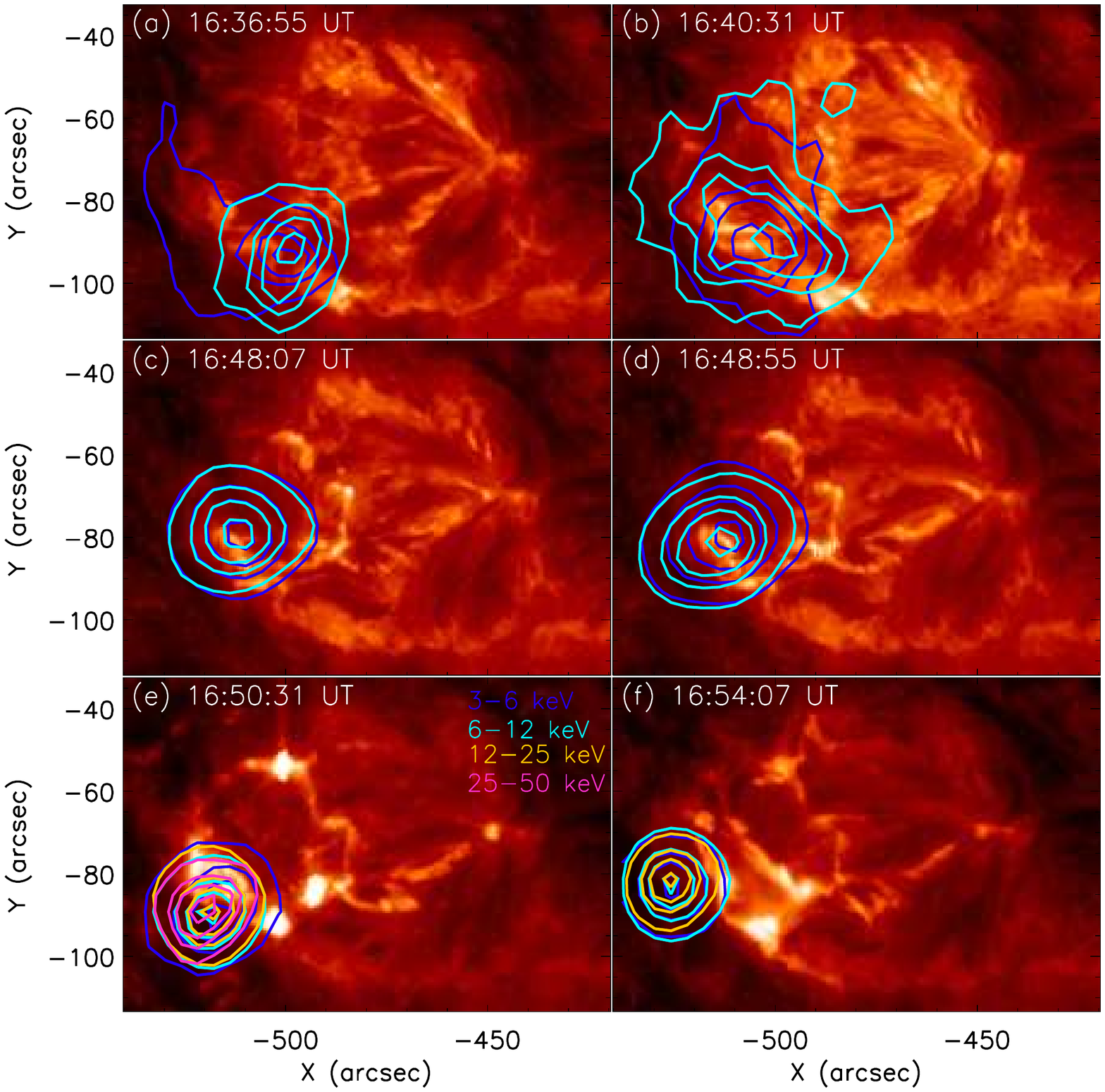}
\caption{Spatial evolution of RHESSI X-ray sources during the pre-flare phase and main flare.
The sources during the pre-flare phase are displayed in panel (a) and (b). From panels (c)-(f), the evolution of
X-ray sources during main flare are shown. Here contours of blue, cyan, yellow, and pink color represent the
X-ray sources of 3--6 keV, 6--12 keV, 12--25 keV, and 25--50 keV energy levels, respectively.
The images of RHESSI are re-constructed by using CLEAN algorithm.
The contour levels are set as 40$\%$, 65$\%$, 80$\%$, and 95$\%$ of the maximum intensity.}
\vspace{-5.0cm}
\label{Fig_rhessi_evolution}
\end{figure*}

The spatial evolution of X-ray sources during the main flare phase is presented in
Figure \ref{Fig_rhessi_evolution}c--f. Up to the impulsive phase of the flare,
 we found X-ray sources only at low
energies (3--6 keV and 6--12 keV). On the other hand, 
during the peak phase, X-ray sources at higher energy bands (12--25 keV and 25--50 keV) were also detected.

\begin{figure*}
\centering
\includegraphics[width=\textwidth]{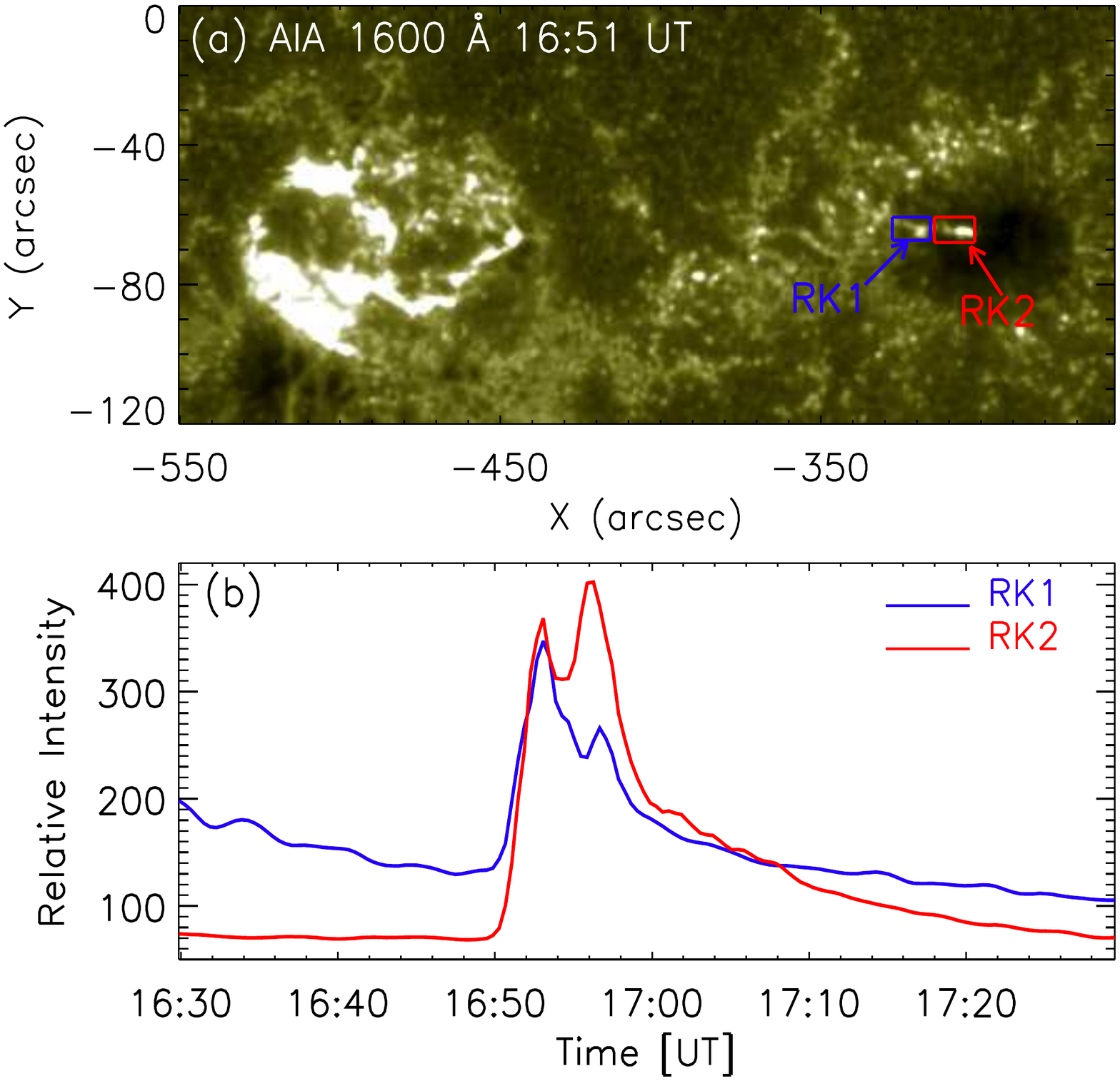}
\caption{AIA 1600 \AA\ image showing the remote ribbons RK1 and RK2 by blue and red boxes,
respectively. The light curve of the two ribbons are displayed in panel (b) showing dual peaks.
The light curves has same colour conventions as in panel (a).}
\vspace{-5cm}
\label{lc1600}
\end{figure*}

\begin{figure*}
\centering
\hspace{-2cm}
\includegraphics[width=\textwidth]{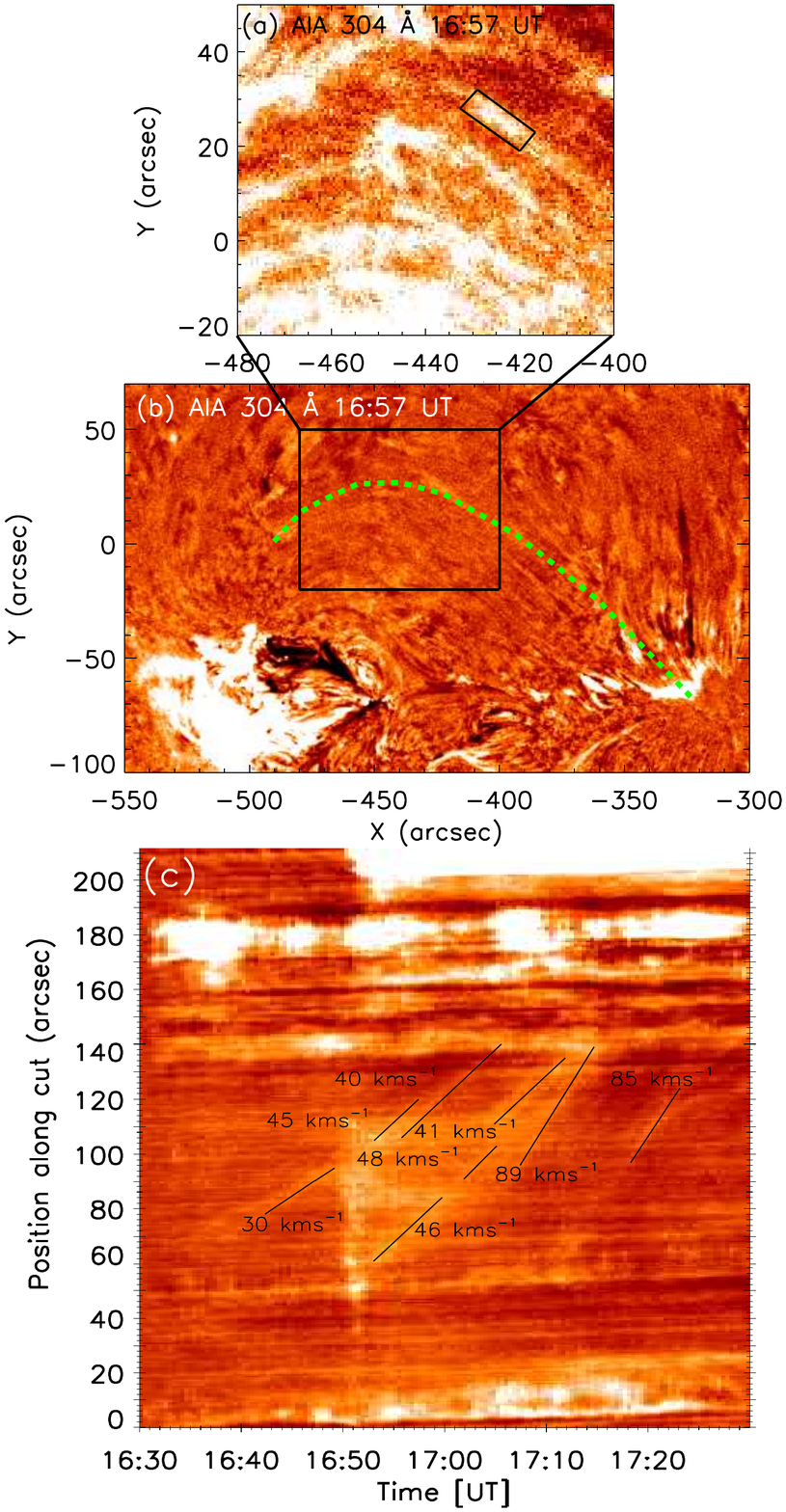}
\caption{AIA 304 \AA\ images showing the slit (green) taken in the direction of motion of the
plasma blob in panel (b). The black rectangle in panel (b) is zoomed in panel (a) which
shows the moving plasma blob by a small black box. Panel (c) displayed the time-distance plot
corresponding to slice shown in (b).}
\label{box_slice_ts}
\end{figure*}
We compare the location of the X-ray sources with AIA/EUV 304 \AA\ images depicted in Figure \ref{Fig_rhessi_evolution}.
The co-aligned images shows that the EUV emission and X-ray sources are co-spatial up to the early impulsive phase of the flare.
During the flare maximum phase ($\approx$ 16:50 UT), the location of X-ray sources is shifted towards south-east
(about 15$''$ towards east and 4$''$ towards south). The shift in the locations of EUV and X-ray sources
could be due to the fact that the sources at different wavelength originate from different heights
 of the solar atmosphere.
Furthermore, RHESSI imaging analysis reveals that both thermal and
non-thermal X-ray sources are co-spatial. The co-spatiality of thermal
and non-thermal sources are not uncommon in compact flares (see {\it e.g.}, \opencite{Kushwaha2014}).
Notably, no distinct foot-point X-ray sources were observed from the locations of
flare ribbons R$_1$and R$_2$.
We also tried to reconstruct the X-ray sources in different energy bands at the site of remote ribbon.
However, the X-ray sources at this site were absent. In this context, it is noteworthy to report two
bright distinct kernels over the remote ribbon in AIA 1600 \AA\ images which are presented in
Figure \ref{lc1600}a (marked as RK1 and RK2).
We present time profiles of intensity of the two kernels in Figure \ref{lc1600}b.
The time profiles indicate a dual peak structure with the first one being co-temporal with the
overall flare maximum observed by GOES 1--8 \AA~channel at $\approx$16:53 UT.
The flare kernels observed in the AIA 1600 \AA~channel indicate that the
source of emission was very compact yet rooted deep up to the upper photospheric levels.

An interesting feature observed in AIA 304 \AA~images is the flow of plasma at EUV
temperatures from the site of circular ribbon to remote ribbon along the overlying field lines.
We present some aspects of these plasma flows in Figure \ref{box_slice_ts}. 
As marked by the black box in the zoomed image of 304 \AA~(Figure \ref{box_slice_ts}a),
the plasma flow was observed in the form of moving blobs of plasma.
In Figure \ref{box_slice_ts}b, we indicate the trajectory followed by the
plasma flow in the 304 \AA\ base difference image by a green dashed line.
A time-distance plot along the selected trajectory is shown in Figure \ref{box_slice_ts}c.
To enhance the bright moving features, the time-distance plot has been constructed by the base
difference images where the AIA image at 16:30:19 UT is adopted as the base image.
The time-distance diagram captures many instances of moving plasma blobs.
However, the flow cannot be observed continuously from the start to the end locations of the chosen
trajectory. The speed of plasma blobs ranges from $\approx$30 \kms to $\approx$90 \kms.

\subsection{Flare Associated Jet Activity}
\begin{figure*}
\centering
\includegraphics[width=\textwidth]{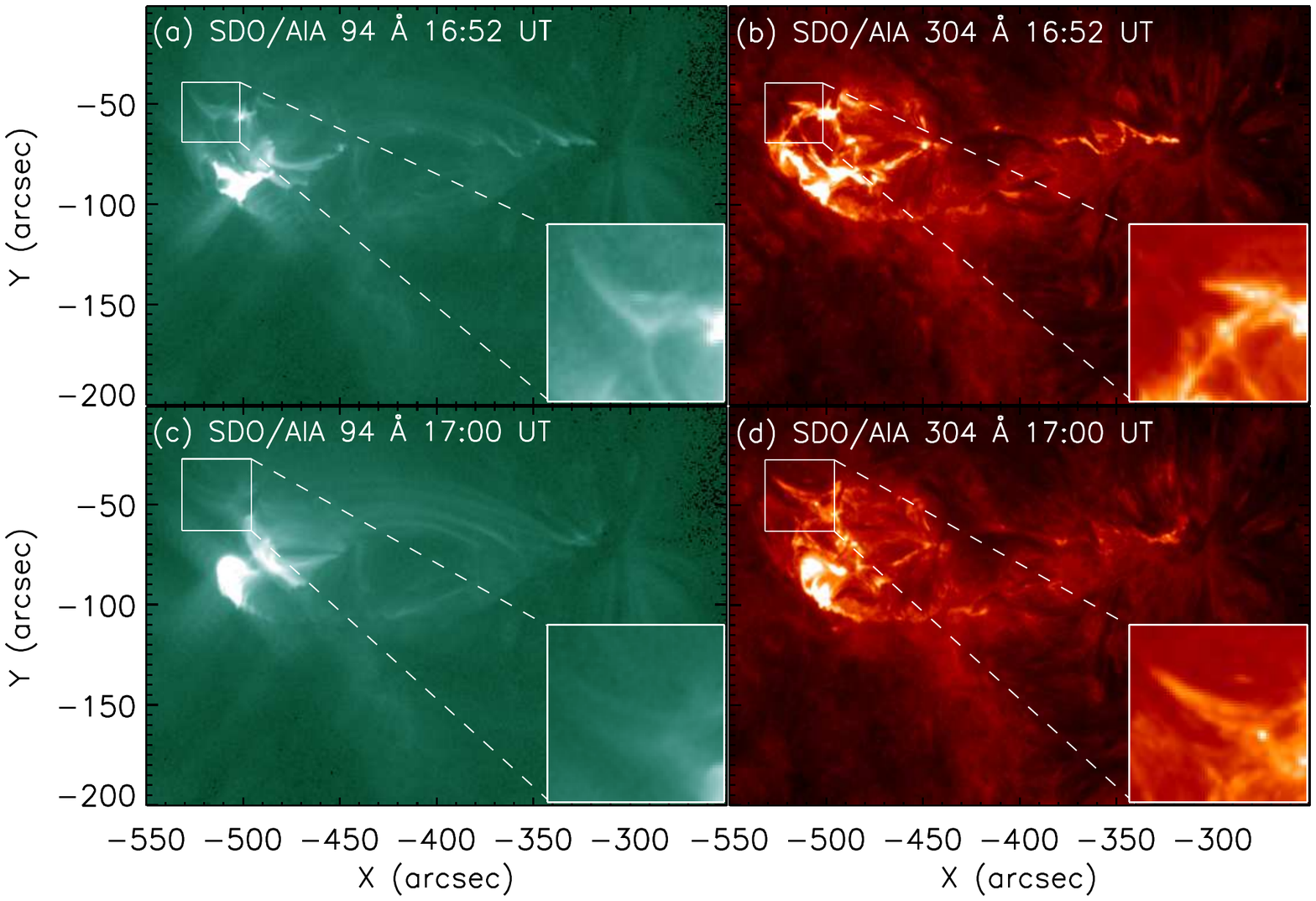}
\caption{SDO/AIA 94 \AA\ and 304 \AA\ images showing the jet activity
associated with the flare region. The inset shows an enlarged view of the jet.}
\vspace{-8.5cm}
\label{Fig_jet}
\end{figure*}

Around 16:52 UT, while the flare was going through the peak phase, jet activity was observed at the north-east
 (-500$''$, -47$''$)
edge of the circular ribbon. 
We have indicated the jet activity in the
94 \AA\ and 304 \AA\ images in Figure \ref{Fig_jet}. 
We notice that in AIA 94 \AA\ wavelength, the jet is visible  
around 16:52 UT, whereas in 304 \AA\ observations, it can be seen at $\approx$17:00 UT.
It means that the jet activity was first seen at higher coronal temperatures 
({\it i.e.} in 94 \AA\ and also in 131 \AA\ wavelength) over the lower temperature emission
that originates 
from transition region and chromospheric layers ({\it i.e.} 304 \AA\ wavelength).
The estimated speed and projected height of the jet are $\approx$300 kms$^{-1}$
and $\approx$30 Mm, respectively, and its life time is $\approx$13 min.
The jet activity continued up to 17:10 UT.
We can see that along the direction of jet eruption the coronal loops 
extended up to large coronal heights.
The association of jet activity with
the circular ribbon flares has been reported in observations and simulations \citep{Pariat10,Wang12,Li18,Zhang18}.

RHESSI X-ray sources up to 50 keV energies were also observed close to the base of the jet activity.
This suggests that the jet material is continuously supplied from the reconnection site.
Along with the jet activity, the loops connecting the positive and negative polarity starts to rise from the
following polarity site, where the circular ribbon was formed. Contrary to this, the other end of the loop system was fixed.
Through these connecting loops, the accelerated non-thermal particles injected from the coronal reconnection region at the
following polarity can travel towards the 
distant foot-points of the leading polarity causing remote
brightenings at that location.
The remote brightening appeared at several places in the eastern part of the leading polarity sunspot.

\subsection{Magnetic Field Evolution and Topology of the Active Region}
\subsubsection{Photospheric Structure}
\begin{figure*}[t]
\centering
\includegraphics[width=\textwidth]{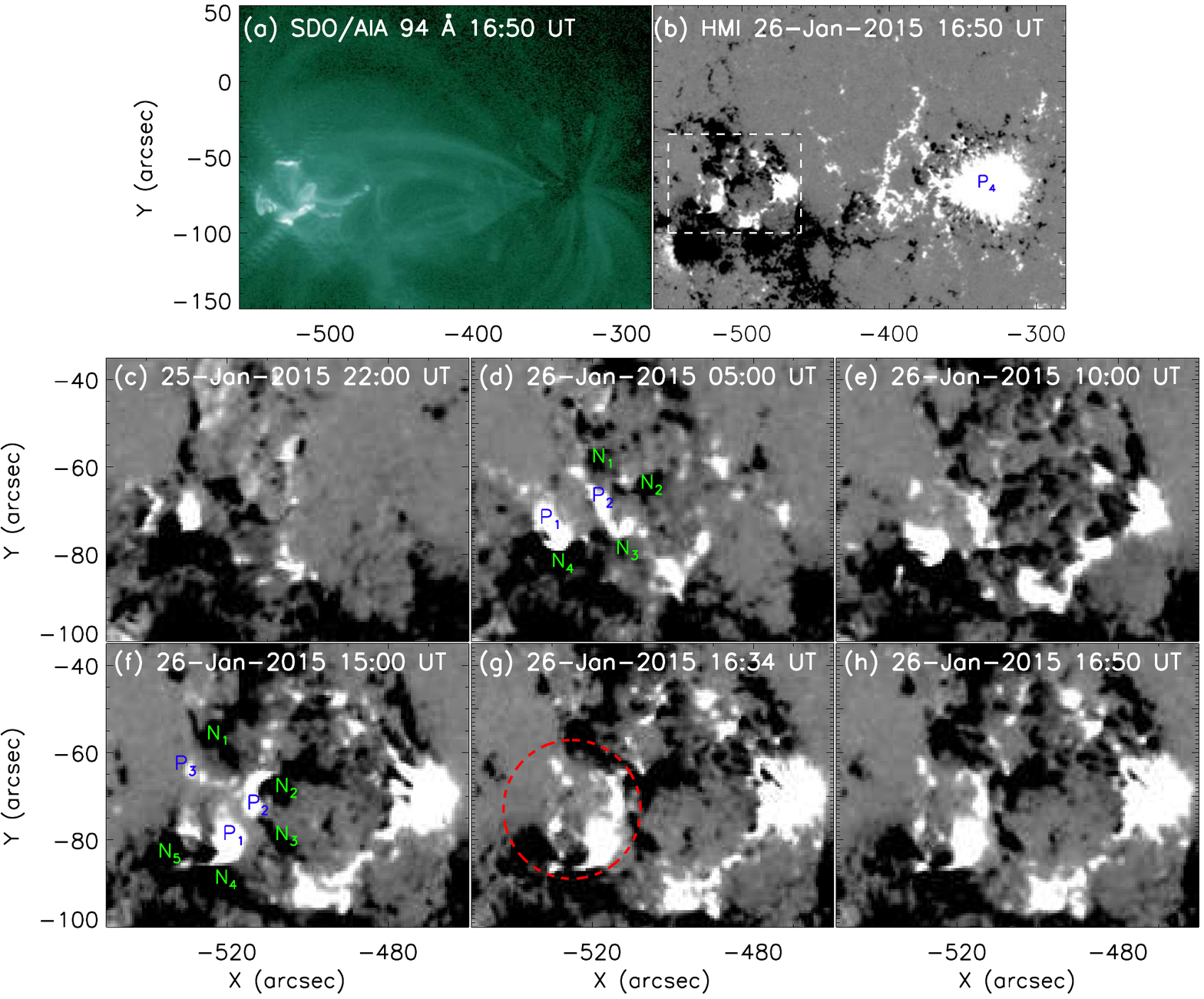}
\caption{Evolution of the magnetic field in the flare region.
Panel (a) shows an AIA 94~\AA~image at 16:50 UT showing the circular ribbon of the flare.
Panel (b) shows an HMI magnetogram of the flare region by dashed box and the remote brightening
region at P$_{4}$. The evolution of the magnetic field for the flaring region with enlarged view of this box is
displayed in panels (c)--(h). Red dashed circle in panel (g) indicate the location of the circular flare ribbon.}
\vspace{-6.8cm}
\label{Fig_hmi}
\end{figure*}

The occurrence of flare and associated phenomena are closely related to the 
active region's magnetic field configuration.
The evolution of the magnetic field is depicted in Figure \ref{Fig_hmi}.
When the active region appeared on the disk, it exhibited mixed
magnetic polarities at both the circular ribbon flare site and the remote ribbon site.
The flaring region (shown with the dashed box in Figure \ref{Fig_hmi}b) shows the
continuous emergence of photospheric magnetic flux of both polarities for a long
period prior to the flare on 26 January 2015. A few of the dispersed patches of the
magnetic field from this region are of special interest, we mark them as P$_{1}$, P$_{2}$, N$_{1}$, N$_{2}$, N$_{3}$,
and N$_{4}$ (shown in Figure \ref{Fig_hmi}d). As time progressed, a part of P$_{1}$ displayed a
slow westward motion and a negative polarity region emerged behind the slowly moving P$_{1}$,
which is indicated as N$_{5}$ in Figure \ref{Fig_hmi}f. Simultaneously, we also observed migration
of a positive polarity patch (indicated as P$_{3}$ in Figure \ref{Fig_hmi}f) towards N$_{1}$.
The positive polarity P$_{1}$ further moved towards P$_{2}$ and finally P$_{1}$ and P$_{2}$ merged around 16:34 UT.
Now, we recognize that at circular ribbon site, the negative polarity patches (N$_{1}$--N$_{5}$)
surrounded the central positive polarity
making the entire arrangement similar to an anemone configuration. This anemone type
photospheric structure was co-spatial with the circular ribbon of the flare,
which is shown by red dashed circle in Figure \ref{Fig_hmi}g.

We also observe continuous emergence and cancellation of the flux at remote
brightening site, where positive polarity is more magnetically dominated.
We name this positive polarity as P$_{4}$ (shown in Figure \ref{Fig_hmi}b).

\subsubsection{Nonlinear Force Free Field (NLFFF) Extrapolation}
\begin{figure*}[t]
\centering
\includegraphics[width=\textwidth]{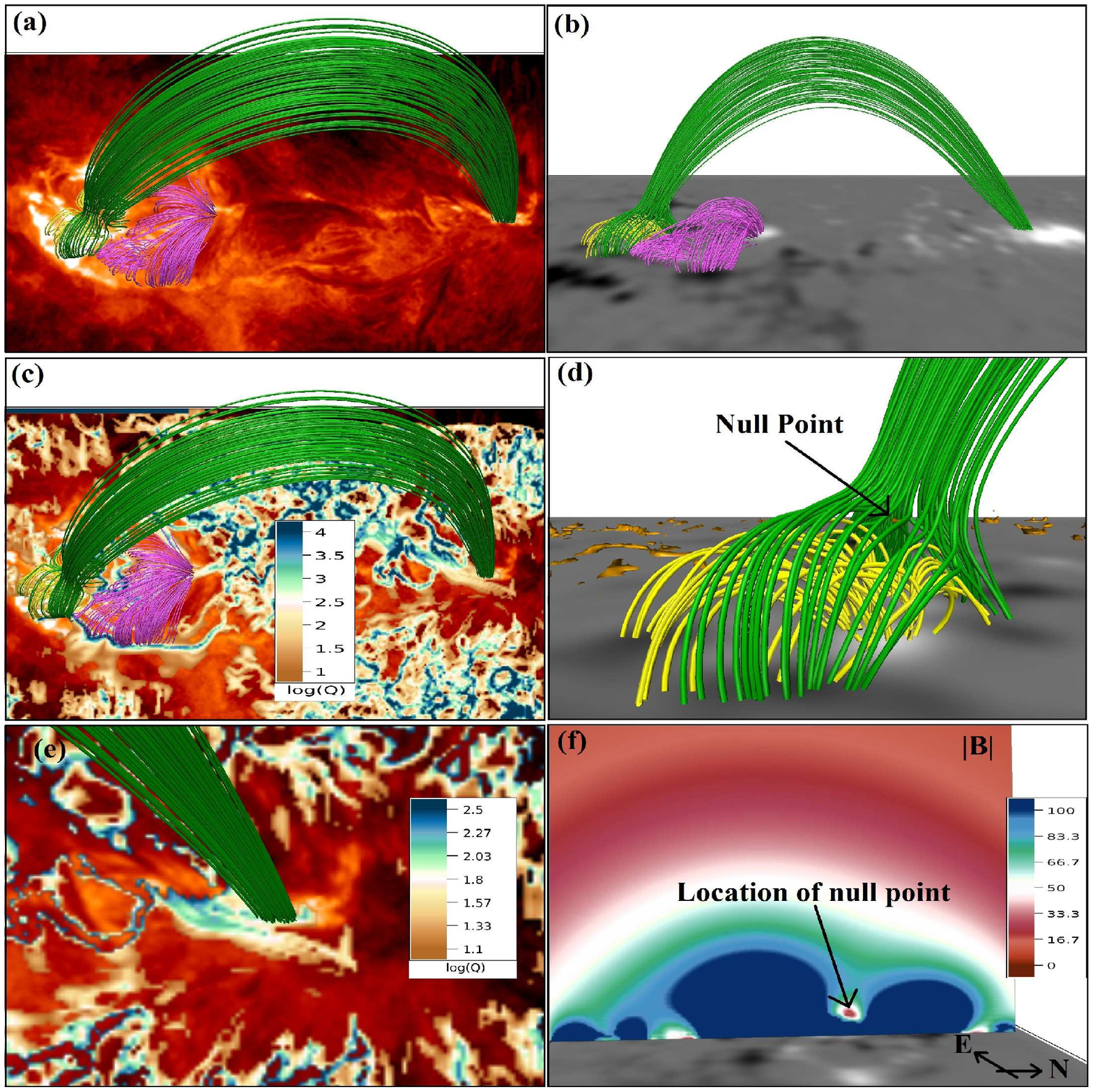}
\caption{Panel (a): Non linear force free field (NLFFF) lines associated with the active region
NOAA 12268 during the pre flare phase. The yellow and green lines form a typical fan-spine
configuration. The circular ribbon region involved the fan-spine configuration along with
the pink lines which are a set of low coronal closed magnetic loops.
For comparison we have plotted an AIA 304 Å image showing the circular ribbons at the base
of the extrapolation volume. Panel (b): NLFFF lines of the overall active region from the
different angle. In panel (c) we plot the degree of squashing factor ($Q$) on the background
AIA 304 \AA. Panel (d): The fan configuration along with a part of the spine lines showing
the approximate location of the null-point (shown by the orange iso-surface).
In panel (e) we plot only the remote ribbon region along with the associated log($Q$) values
between [1.1, 2.5]. Panel (f): A vertical surface passing through the null-point showing
the variation of magnetic field strength ($|$B$|$). Note that, the surface in panel (d)
is along Y--Z orientation and we have indicated the north and east directions in the same
panel for convenience.}
\vspace{-5cm}
\label{Fig:NLFFF}
\end{figure*}

To investigate the coronal magnetic configuration
associated with the active region,
we have performed a non-linear force free field (NLFFF) 
extrapolation using the advanced version of an optimization 
code developed by \citet{Wiegelmann07} and  \citet{Wiegelmann10}. 
In Figure \ref{Fig:NLFFF}, we have displayed a few of the modelled 
field lines representing the coronal connectivities 
prior to the flare to show the coronal magnetic field
configuration of the active region from different viewing angles.
The NLFFF extrapolation result 
suggests the presence of a fan-spine configuration
involving a coronal null-point in the active region.
Notably, the fan lines (shown by yellow colour in Figure \ref{Fig:NLFFF}a--d)
originated from the mixed polarity trailing region which was
associated with the circular ribbon brightening (highlighted by
the dashed circle in Figure \ref{Fig_hmi}g). This mixed polarity
region region was also connected with the leading positive polarity
sunspot (indicated by P$_{4}$ in Figure \ref{Fig_hmi}b), by a set
of large closed lines (spine lines).\

As described in sub-section 3.3 and displayed in AIA 94 \AA\ image of Figure \ref{Fig_mosac94},
we have found that the main circular ribbon flare and the jet site is connected
to the remote flare site through high-lying coronal loops. The jet material ejects along these
high-lying coronal loops. We have modelled these loops using the
NLFFF extrapolation and found two type of field lines (shown in Figure \ref{Fig:NLFFF}):
first are yellow low-lying field lines (fan-structure) and the second are 
green high-lying field lines (spine-structure).
Usually the jets are observed at the location of open magnetic field lines and newly emerged magnetic flux, which
formed small field lines (see {\it e.g.}, \opencite{Chandra17,JoshiR2017} and references cited therein). 
However, in our case we found that a set of large-overlying field line originate from
the jet location which closes at larger distances. Thus, it is likely that the coronal
jet resulted as the low-lying bipolar loops within the circular flare area interacted
with the large, apparently uni-polar field lines essentially forming leg of a high-lying
coronal loop system. Therefore, our present case evidenced that the observed jet is due
to the reconnection between the low-lying and the high-lying magnetic field lines.
Such a scenario was also reported in previous observations \citep{Bentley2000,Uddin2012}. 
In such type of confined magnetic configuration, the material ejected from the jet and
associated accelerated particles are expected to largely remain confined in closed high-lying loops.

In order to find the photospheric regions with high connectivity gradient,
we calculated the degree of squashing factor ($Q$) by employing the code developed by \inlinecite{Liu2016} 
based on the NLFFF extrapolation results. From Figure \ref{Fig:NLFFF}c, we readily find that the circular
ribbon region was characterized by the highest $Q$ values in the active region (log($Q$)$>$3.5).
Further it is also noticeable that, the regions associated with the remote ribbon had moderately
higher values of $Q$ compared to the ribbon-less regions. From Figure \ref{Fig:NLFFF}e,
we find that the remote brightening region was associated with $Q$ values between
$\approx$30 and 200 (1.5$>$log($Q$)$>$2.3).

\section{Discussion and Conclusion}
In this article, we investigate an M1.1 class solar flare
of 26 January 2015. An important finding of this study lies in the complex morphological
evolution of flare ribbon in anemone-type magnetic field configuration within active region
NOAA 12268, where this confined M1.1 class flare occurred. The flare rises with the development of the
circular ribbon which eventually proceeds with three parallel ribbon structures within the
same region. The main results of our study are summarized as below:
\begin{enumerate}
\renewcommand{\labelenumi}{\roman{enumi})}
\item
	{We have observed the pre-flare activity in
the active region that spans for $\approx$15 min in EUV/UV as well as in 
X-ray energy bands. This pre-flare activity could provide an important
role for the triggering of main circular ribbon flare.}

\
\item 
{The evolution of loops visible in AIA 193 \AA\ and 94 \AA\ reveals following connectivity:
	We observed the loops originating from the circular ribbon region connect
	and converge to the distant positive polarity sunspot region such that the overall
	structure resembles with fan-spine configuration.
In the decay phase of flare, where two ribbons appeared at the
south-east site, we observed the post-flare loops as visible in a typical two ribbon flares.
NLFFF extrapolation of the active region evidenced the fan-spine 
magnetic topology and the presence of null-point in the active region.}

\
\item
{The remote flare ribbon exhibited fine structured kernels.
An important characteristic of remote ribbon is the progression of brightness
from east to west direction. Further, photospheric region corresponding to the remote
ribbon indicates relatively higher values of degree of squashing factor ($Q$) that
lie in the range of $\approx$30 and 200, providing evidence of  slipping reconnection.}
\end{enumerate}

We have observed the pre-flare activity in the active region prior
to $\approx$15 min of the main flare.
The pre-flare activity in solar flares were also reported in the past 
literature \citep{Veronig2002,Kim07,Chandra09,Joshi11,AwasthiMNRAS14,Wang17,Hernandez-Perez2019,Mitra2019,Mitra2020}.
In our study, the co-spatiality of the pre-flare and main flare emitting regions in X-ray 
and EUV/UV provides evidence that pre-flare activity
has likely contributed toward triggering of the main flare.
In this context, it is noteworthy that in terms of morphology of hot plasma
structures the pre-flare and the main flare present similarities.
The regions of pre-flare brightenings continued to be visible during
the main flare, albeit with variation in intensity ({\it cf.} Figure \ref{Fig_mosac304}b and \ref{Fig_mosac304}d).

To find a plausible interpretation of the observed pre-flare activity,
we examine the corresponding region in the photospheric magnetograms during the period
of pre-flare activity (Figure \ref{Fig_hmi}g).
By this time, the photospheric magnetic field configuration of the flaring region
had evolved into an $``$anemone$"$ type configuration (see the region delineated by
dashed circle in Figure \ref{Fig_hmi}g).
The NLFFF extrapolation suggests a fan-spine configuration involving a coronal null-point
with outer fan-lines being anchored at the periphery of the photospheric anemone structure
(Figure \ref{Fig:NLFFF}a--d).
These observations indicate that the pre-flare brightening, occurring within the periphery
of later developed circular ribbon, is suggestive of the ongoing null-point reconnection
which is quite subtle at this time with no detectable emission at the location of remote ribbons.
We further emphasize that the pre-flare magnetic configuration, as discussed above,
has evolved as a result of continuous flux emergence and cancellation (see Figure \ref{Fig_hmi}
and section 3.4.1).
In such regions, the pre-flare brightenings are also attributed to low-atmosphere magnetic
reconnection ({\it e.g.} \opencite{Moon2004,Wang17}).
Thus our observations suggest the contribution from both the processes described above
that eventually led to a mild yet continuous energy release $\approx$15 min prior to the
main circular ribbon flare.
\opencite{Green2018} presents a review of the predictability of solar eruptions.
They have discussed that the numerical simulations and observations (see their table 1)
provide evidences that the pre-flare configuration play an important role for the
major flare activity. The pre-flare activity lower the magnetic tension
in the active region and as a result of this the active region produced main flare.

The most noticeable characteristic of the M1.1 flare is the development of a circular ribbon
structure and brightening at a distant location which eventually progresses into the remote
ribbon (Figure \ref{Fig_304light_curve}). The circular ribbon is visible in almost all AIA EUV/UV filters.
The circular ribbon structures and associated remote brightening can be explained by magnetic
reconnection at a coronal null-point located in the fan-spine configuration
\citep{Masson09,Reid12,Wang12,Pontin2016,Hao17,Hernandez-Perez2017,Li18,Zhang18}.
The topology of the flaring region, inferred from the NLFFF extrapolation, readily supports this scenario
(Figure \ref{Fig:NLFFF}a--c) where we find a good match between observed and modelled structures.
These observed structures include the circular ribbon, remote ribbon, and the overlying loop system
connecting the circular and remote ribbons. We observed a time delay of $\approx$2 min in the appearance
of remote brightenings with respect to the circular ribbon appearance (Figure \ref{Fig_304light_curve}).
The time delay in the evolution of the circular and remote ribbons has been well documented in
previous studies ({\it e.g.} \opencite{Wang12,Li18}).

During the main phase of the M-class flare, we notice moving blobs of plasma that apparently
follow the overlying coronal loops connecting the circular and remote ribbons.
The time-distance diagram based on base difference images (Figure \ref{box_slice_ts})
clearly indicates these plasma flows but they occur within a limited range of distances only.
\inlinecite{Hernandez-Perez2017} reported plasma flows toward the formation site of remote ribbons from
the location of the circular ribbon in the flare of 29 January 2015, which occurred in the same active region.
However, unlike our case, the plasma flows in their event occurred during the pre-flare phase
(as early as 15 min before the onset) and the continuous flow patterns were visible between
the two distant ribbon locations. \inlinecite{Hernandez-Perez2017}, therefore, concluded that the
remote brightenings were caused by the dissipation of kinetic energy of the plasma flows.
In our observations, the moving plasma blobs appear as temporary structures that fade away
much before reaching to the location of the remote ribbon.
Therefore, it is more likely that the remote ribbon in our case is primarily formed by the interaction
of non-thermal particles, flowing along the overlying coronal loop system connecting the location
of circular ribbon to its remote footpoint, which is well observed in AIA 94 \AA\ images
(see the movie accompanying Figure \ref{Fig_mosac94}).
Our interpretation is consistent with the mechanism of typical circular ribbon flares
proposed in some of the earlier studies ({\it e.g.} \opencite{Tang1982,Nakajima1985,Wang12}).
The streams of accelerated particles will eventually be collisionally
stopped by the denser layers of the lower solar atmosphere creating the fine-structured
kernels of the remote ribbon.

A careful examination of the remote ribbon suggests the spatial progression of brightening from east to west direction.
In this context, from Figure \ref{Fig:NLFFF}e, we note that the remote brightening region is associated
with high degree of squashing factor ($Q$).
The calculated value of this $Q$ is found in between 30 and 200.
From such high values of $Q$, the motion of the remote brightening can be interpreted in terms of slipping reconnection
in this region \citep{Chandra11,Aulanier12,Dudik14,Li2014,Li2015,Janvier15,Dudik2016,Zheng2016}.
In fan-spine topology, the slipping magnetic reconnection in QSLs and null-point
reconnection can occur simultaneously \citep{Masson09,Reid12,Pontin2016}.

Before the main flare, significant amount of the positive magnetic
flux emerged in the active region (Figure \ref{Fig_hmi}).
This positive polarity flux emergence got distributed
within the pre-existing patches of negative polarity in such a way
that an overall anemone-type magnetic configuration appeared in the
photosphere \citep{Asai08}. This kind of magnetic flux distribution
in the photosphere is favorable for the fan-spine structure in the corona.\

The NLFFF extrapolation reveals the presence of 3D null at a relatively
lower height {\it i.e.} 10$\pm$1 Mm (Figure \ref{Fig:NLFFF}f). The appearance
of flare brightening in a quasi-circular pattern right from pre-flare
phase, therefore, suggests energy release due to ongoing 3D reconnection
which get enhanced during flare's main phase. The formation of parallel
ribbons within the periphery of the circular ribbon suggests onset of
standard flare reconnection in the later phase of M1.1 flare.
The formation of parallel ribbon in a confined flare is rather surprising
and may point toward the small scale flux rope eruption (invisible in
direct observation and extrapolation) or interaction between different
flux systems within the fan-dome. These complex observations put constraints
on contemporary models of solar flare.
The observations of jet activity in the main flare site also
support the idea that the magnetic reconnection was 
initiated at a coronal null-point \citep{Pariat10, Liu2011, Zeng2016}.

\begin{acks}
We would like to thank the referee for valuable comments and suggestions.
SDO is a mission for NASA's Living With a
Star (LWS) Program. SDO data are courtesy of the NASA/SDO and HMI science team. RHESSI is a NASA Small Explorer Mission.  GOES is a joint effort of NASA and the National Oceanic and Atmospheric Administration (NOAA).
PD, RC and BJ acknowledges the support from the DST/SERB project.
PD also acknowledges the support from the CSIR, New Delhi.
This work is supported by the Indo-Austrian joint research project no. INT/AUSTRIA/BMWF/ P-05/2017 and
OeAD project no. IN 03/2017.
A.M.V. acknowledges the Austrian Science Fund ( FWF) : P25383-N27.
RJ thanks to the Department of Science and Technology (DST), New Delhi, India for 
an INSPIRE fellowship.
\vspace{3mm}

\noindent {\bf Disclosure of Potential Conflicts of Interest}  The authors declare that they have no conflicts of interest.
\end{acks}
\mbox{}~\\
\bibliographystyle{spr-mp-sola}
\bibliography{reference}
\IfFileExists{\jobname.bbl}{} {\typeout{}
\typeout{***************************************************************}
\typeout{***************************************************************}
\typeout{** Please run "bibtex \jobname" to obtain the bibliography}
\typeout{** and re-run "latex \jobname" twice to fix references}
\typeout{***************************************************************}
\typeout{***************************************************************}
\typeout{}}

\end{article} 
\end{document}